\documentclass[prb,twocolumn,preprintnumbers,amsmath,amssymb,superscriptaddress]{revtex4}
 
\usepackage{graphicx}
\usepackage{dcolumn}
\usepackage{bm}
\usepackage{color}

\begin{document}

\preprint{draft}

\title{
Coherent Manipulation of Individual Electron Spin in a Double Quantum Dot Integrated with a Micro-Magnet}

\author{Toshiaki Obata}
\affiliation{%
Quantum Spin Information Project, ICORP,JST,
Atsugi-shi, Kanagawa 243-0198, Japan
}%

\author{Michel Pioro-Ladri{\`e}re}
\affiliation{%
Quantum Spin Information Project, ICORP,JST,
Atsugi-shi, Kanagawa 243-0198, Japan
}%
\affiliation{Department de Physique, Universit{\'e} de Sherbrooke, Sherbrooke, Quebec, Canada J1K-2R1}

\author{Yasuhiro Tokura}
\affiliation{%
Quantum Spin Information Project, ICORP,JST,
Atsugi-shi, Kanagawa 243-0198, Japan
}%
\affiliation{
NTT Basic Research Laboratories, NTT Corporation, 
Atsugi-shi, Kanagawa 243-0198, Japan
}
\author{Yun-Sok Shin}
\author{Toshihiro Kubo}
\author{Katsuharu Yoshida}
\affiliation{%
Quantum Spin Information Project, ICORP,JST,
Atsugi-shi, Kanagawa 243-0198, Japan
}%
\author{Tomoyasu Taniyama}
\affiliation{
PRESTO, Japan Science and Technology Agency (JST), Honcho 4-1-8, Kawaguchi, Saitama 332-0012, Japan
}
\affiliation{
Materials and Structures Laboratory, Tokyo Institute of Technology, Nagatsuta 4259, Yokohama, Japan
}
\author{Seigo Tarucha}
\affiliation{%
Quantum Spin Information Project, ICORP,JST,
Atsugi-shi, Kanagawa 243-0198, Japan
}%
\affiliation{%
Institute for Nano Quantum Information Electronics, University of Tokyo, Komaba, Meguro-ku, Tokyo, Japan
}%
\affiliation{%
QPEC and Department of Applied Physics, University of Tokyo, Bunkyo-ku, Tokyo 113-8656, Japan
}%

\date{\today}

\begin{abstract}
We report the coherent manipulation of electron spins in a double quantum dot integrated with a micro-magnet. We performed electric dipole spin resonance experiments in the continuous wave (CW) and pump-and-probe modes. We observed two resonant CW peaks and two Rabi oscillations of the quantum dot current by sweeping an external magnetic field at a fixed frequency. 
Two peaks and oscillations are measured at different resonant magnetic field, which reflects the fact that the local magnetic fields at each quantum dot are modulated by the stray field of a micro-magnet. 
As predicted with a density matrix approach, the CW current is quadratic with respect to microwave (MW) voltage while the Rabi frequency ($\nu_{\rm Rabi}$) is linear. The difference between the $\nu_{\rm Rabi}$ values of two Rabi oscillations directly reflects the MW electric field across the two dots. These results show that the spins on each dot can be manipulated coherently at will by tuning the micro-magnet alignment and MW electric field.
\end{abstract}

\pacs{73.63.Kv, 03.67.Lx, 75.75.-c, 76.30.-v}
%

\maketitle


\section{INTRODUCTION}

Following Loss and DiVincenzo's proposal of spin-based quantum computing with quantum dots (QDs),\cite{Loss} and motivated by the realization of well-defined single and double QDs,\cite{Tarucha,Ciorga,Johnson,Delft,Hatano} and the observation of a robust spin degree of freedom,\cite{Fujisawa, Elzerman, Petta, Amasha} considerable effort has been devoted to implementing electron spin qubits with QDs. The requirements are the coherent manipulation and detection of single electronic spins, and both have recently been met with an electron spin resonance (ESR) technique \cite{Koppens, Nowack, Laird, Michel} for a double QD in a Pauli spin blockade condition.\cite{Ono, Johnson, Koppens} 
The application of electron spin resonance to spin qubits still presented a challenge, because a sufficiently strong ac magnetic field has to operate on single electrons in a QD. A straightforward technique was initially developed, which involves using a micro-coil placed on top of a QD with an ac current flowing through it.\cite{Koppens} 
However, this is accompanied by joule heating caused by the mA order current flowing through the coil, and so is not useful for making multiple qubits. Electric dipole induced spin resonance (EDSR) is a way to avoid such joule heating, and has been demonstrated using a spin-orbit interaction \cite{Nowack,Golovach} and an inhomogeneous hyperfine field.\cite{Laird} With both techniques, a local ac magnetic field for a QD is generated by employing a microwave (MW) electric field to the QD. More recently, we proposed and demonstrated a technique using a slanting Zeeman field imposed by a micro-magnet,\cite{Tokura, MichelAPL, Michel} which produces a stray magnetic field across a QD. The transverse component is a magnetic field gradient of $\sim $ T/${\rm \mu}$m perpendicular to the externally applied dc magnetic field, and a local effective MW magnetic field is generated by applying an MW electric field to oscillate an electron inside the dot. The longitudinal component is an inhomogeneous magnetic field parallel to the external field, which weakly modulates Zeeman energy across two dots. This field component depends on the micro-magnet geometry relative to the QD, and therefore can be used to selectively address two or more electrons in a coupled multiple QD at different spin resonance frequencies, leading to scalable qubits with a QD array.\cite{Michel,MichelNJP} 

Double QDs with two electrons are basic elements for operating various quantum gates, such as swap and control-not gates,\cite{Petta} all of which utilize the rotation of individual electron spins and the modulation of exchange coupling between electrons. The manipulation of two-electron spins has been demonstrated for swap by electrically modulating the inter-dot tunnel coupling in a double QD (Ref. \onlinecite{Petta}) but not the coherent selective manipulation of individual electron spins or two spin qubits. 
In this paper, we report the coherent manipulation of individual electron spins in a series coupled double QD using EDSR combined with the micro-magnet effect. We measure a current flowing through a double QD to detect the EDSR current and Rabi oscillations in response to MW irradiation of the double dot in the continuous wave (CW) mode and pump-and-probe (p-p) mode, respectively. The CW EDSR peak and the Rabi frequency measured for various MW powers are both higher for one of the two dots located closer to the MW gate, reflecting the effect of the larger MW electric field on this dot. In addition, they increase quadratically and linearly, respectively, with MW voltage. 
These results coincide consistently with the values calculated for our EDSR scheme using the density matrix approach.\cite{Tokura, MichelAPL} 
We present the results of the CW and pump-and-probe experiments in Secs. II and III, respectively. We compare these results quantitatively in Sec. IV in terms of the MW electric field distribution calculated using photon assisted tunneling spectroscopy, and we provide our conclusion in Sec. V. 

\section{CONTINUOUS WAVE EXPERIMENT}
\begin{figure}
\begin{minipage}{0.22\textwidth}
\begin{flushleft}
\includegraphics[width=\textwidth]{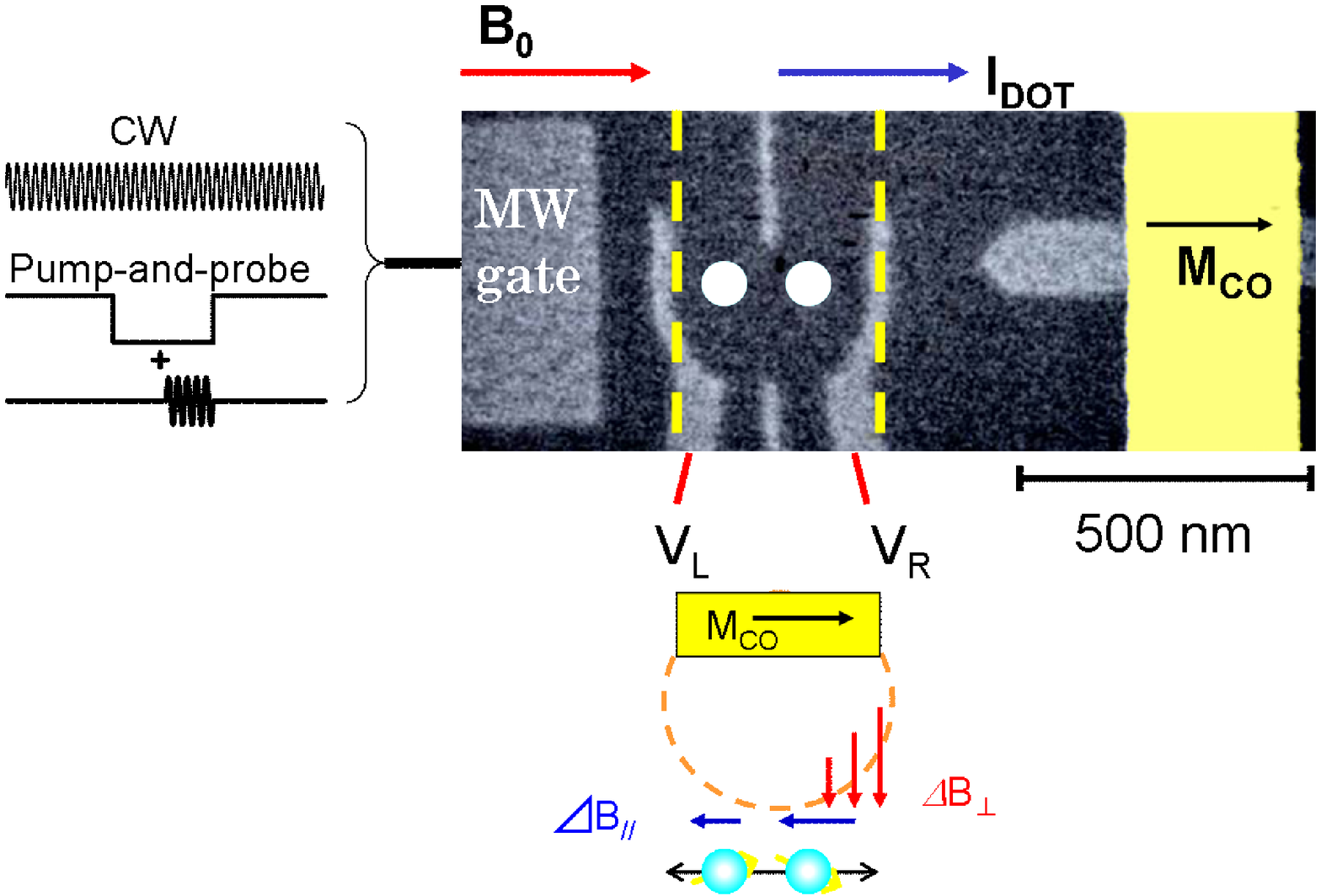}\\ 
(a)
\end{flushleft}
\end{minipage}
\begin{minipage}{0.23\textwidth}
\begin{flushleft}
\includegraphics[width=\textwidth]{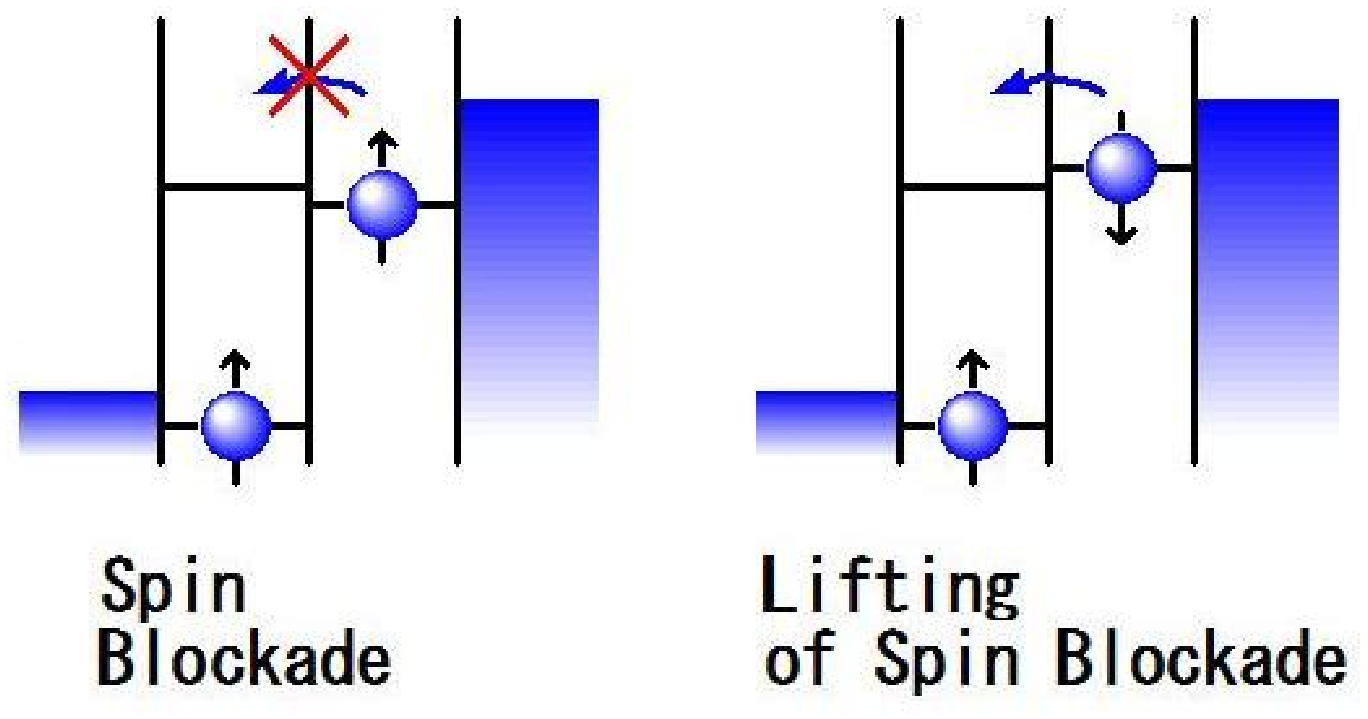}\\ 
(b)
\end{flushleft}
\end{minipage}
\caption{\label{fig:sampleSEM} 
(a) Scanning electron micrograph of the device. 
Two electrons are confined in a lateral double quantum dot by Schottky gates (gray) deposited on the surface of an AlGaAs/GaAs hetero-structure, in which a two-dimensional electron gas is located 100 nm below the surface. An MW gate is placed to the left of the double dot. A cobalt micro-magnet (shown in yellow/light gray) is placed on top of the dot with a 100 nm thick calixarene insulator film between them. 
In this image, the cobalt micro-magnet is intentionally displaced to show the arrangement of the Schottky gate electrodes. 
The device is placed in a dilution refrigerator with an electron temperature of $\sim$ 150 mK that was estimated by measuring the peak width of the Coulomb peak. 
(b) Energy diagram of the double quantum dot used for the CW EDSR experiment. 
}
\end{figure}
We micro-fabricated a lateral dot sample by using electron beam and optical lithography as well as vacuum electron beam deposition.
Figure \ref{fig:sampleSEM}(a) shows a scanning electron micrograph of a device that we made for a fabrication test. It is similar to the device used in our experiment. We isolate two electrons and separate them from each other spatially using a gate-defined double QD,\cite{Michel} connected in series to source and drain reservoirs. The QD parameters estimated from the measurement of stability diagrams are as follows; tunnel coupling t= 0.83 $\rm \mu$eV, nominal charging energy, V$_{\rm intra}$ = 5.0 meV, and inter-dot coupling energy, V$_{\rm inter}$ = 1 meV. 
The micro-magnet is magnetized in-plane (the magnetization is shown by an arrow labeled M$\rm_{Co}$) by applying a sufficiently large external in-plane magnetic field $B_0$ exceeding 0.5 T.\cite{Michel} The micro-magnet is located above the double dot, and generates static out-of-plane (red arrows) and in-plane (blue arrows) stray fields. In our calculation, the out-of-plane stray field has a large gradient of 0.8 T/$\rm \mu$m, and the in-plane fields at the two dots differ by 20 $\sim$ 30 mT. The MW electric field, $E_{\rm MW}$, is produced by applying an MW voltage, $V_{\rm MW}$, to the MW gate electrode in Fig. \ref{fig:sampleSEM}(a) closer to the left QD of the two QDs. The same electrode is used to detune the inter-dot energy levels in the p-p measurements. 

\begin{figure}
\begin{minipage}{0.25\textwidth}
\begin{flushleft}
\includegraphics[width=\textwidth]{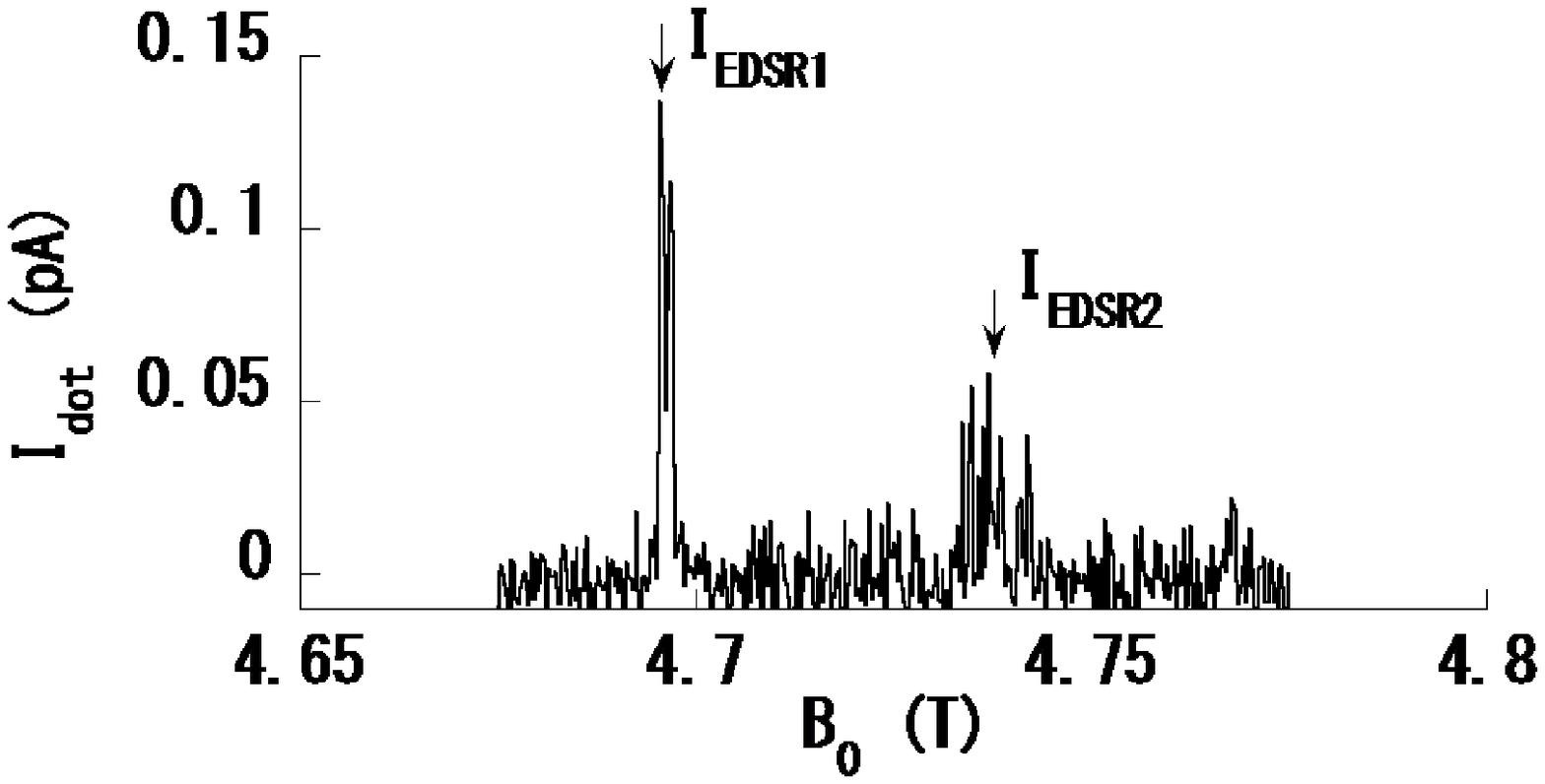}\\ 
(a)
\end{flushleft}
\end{minipage}
\begin{minipage}{0.20\textwidth}
\begin{flushleft}
\includegraphics[width=\textwidth]{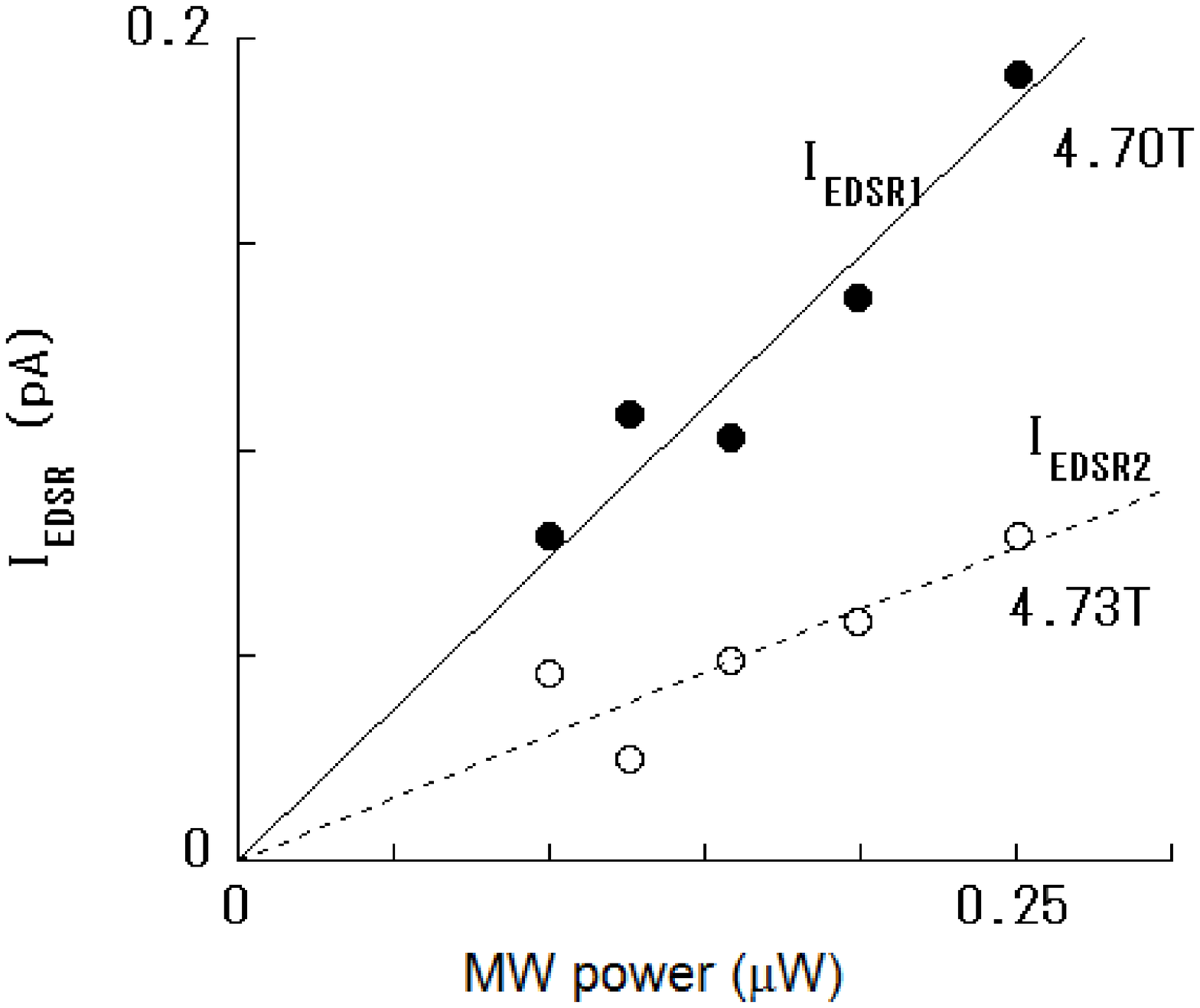}\\ 
(b)
\end{flushleft}
\end{minipage}
\caption{\label{fig:CW} 
(a) Current I$_{\rm dot}$ as a function of external magnetic field $B_0$ for MW irradiation at a frequency of 25.6 GHz and a power of -37 dBm. High and low EDSR peaks are observed at 4.70 and 4.73 T, respectively. 
(b) EDSR current $I_{\rm EDSR1,2}$ vs MW input power. The local electric field depends on the position of each dot, and is proportional to the square root of the MW power. 
The slopes of the two straight lines are quadratic to $E_{\rm EDSR1}$ and $E_{\rm EDSR2}$, 
and the ratio of the two peaks $I_{\rm EDSR1}$ and $I_{\rm EDSR2}$, $(E_{\rm EDSR1}/E_{\rm EDSR2})^2$, agrees with the values evaluated in the following discussions.
}
\end{figure}

To detect EDSR, we placed the double dot in a Pauli spin blockade (P-SB) regime in Fig. \ref{fig:sampleSEM}(b) and applied a CW MW. Figure \ref{fig:CW}(a) shows the current $I_{\rm dot}$ through the double dot vs the external magnetic field $B_0$ measured for the MW at 25.6 GHz (note that we previously obtained similar data using the same device but under different conditions.). The source-drain bias was set at 1.0 $\sim$ 1.5 mV, which is at least 10 times larger than the single photon energy (typically less than 100 $\rm \mu$eV), to avoid photon assisted tunneling (PAT) through the outer barriers. In this figure, we can see two peaks separated by 30 mT, reflecting the EDSR for the individual electron spins in the two dots. The current through the double dot is initially blocked by Pauli exclusion once a spin triplet state has been formed in the double dot.\cite{Ono} The spin-flip induced by EDSR dissolves the stacked state of spin triplet states. In our scheme, only one of two electrons spins flips on resonance while the other stays the same in the off-resonance state, and the electron state transits to the mixed state of a spin triplet and singlet state, which lifts the blockade and gives rise to a finite leakage current in the resonance condition on either dot. 

With low MW power, the EDSR peak height, $I_{\rm EDSR}$, is proportional to the square of the MW induced magnetic field, $B_{\rm MW}$.
By using a standard density matrix approach, we can explicitly derive the formula $I_{\rm EDSR} \cong \frac{2 \pi ^2 e \Gamma_\phi \nu_{\rm Rabi}^2}{(\Gamma_\phi)^2+\delta^2}$ for the lowest order of $\nu _{\rm Rabi}$ where $\nu_{\rm Rabi}$ is the Rabi frequency proportional to $B_{\rm MW}$, therefore $I_{\rm EDSR}$ is quadratic with respect to the MW induced magnetic field, $B_{\rm MW}$.
$\Gamma_\phi$ and $\delta$ are the decoherence rate and the inter-dot energy detuning, respectively. 
The components of $\Gamma_\phi$ are the spin decoherence rate and the inter-dot inelastic tunneling rate. 
$B_{\rm MW}$ is proportional to the field gradient times the root-mean-square displacement of an electron and, therefore, $I_{\rm EDSR} \propto (E_{{\rm MW}_i} b_{{\rm SL}_i})^2$, where $E_{{\rm MW}_i}$ and $b_{{\rm SL}_i}$ (i=1,2) are the amplitude of the MW electric field and the out-of-plane magnetic field gradient across the double dot, respectively.\cite{Michel} Here, i = 1 and 2 for the left and right QDs, respectively. Since $b_{\rm SL}$ is almost uniform over the double dot, the higher amplitude of the peak at a smaller $B_0$ in Fig. \ref{fig:CW}(a) is assigned to the EDSR for the spin located in the left QD closer to the MW gate, because $E_{{\rm MW}_1}>E_{{\rm MW}_2}$ and thus $B_{{\rm MW}_1}>B_{{\rm MW}_2}$. 
We plot the MW power dependence of each EDSR peak height $I_{\rm EDSR}$ in Fig. \ref{fig:CW}(b). 
The power dependences for each peak fit well to a linear relation. The MW power is quadratic with respect to the magnitude of the MW electric field, and this indicates that $I_{\rm EDSR} \propto E_{\rm MW}^2$ as expected. So from the ratio of the slopes between the two straight lines, we are able to calculate an $E_{\rm MW}$ ratio of 1.58 (= $\sqrt{2.5}$) between the two dots.

\section{PUMP-AND-PROBE EXPERIMENT}

\begin{figure}
\begin{minipage}{0.225\textwidth}
\begin{flushleft}
\includegraphics[width=\textwidth]{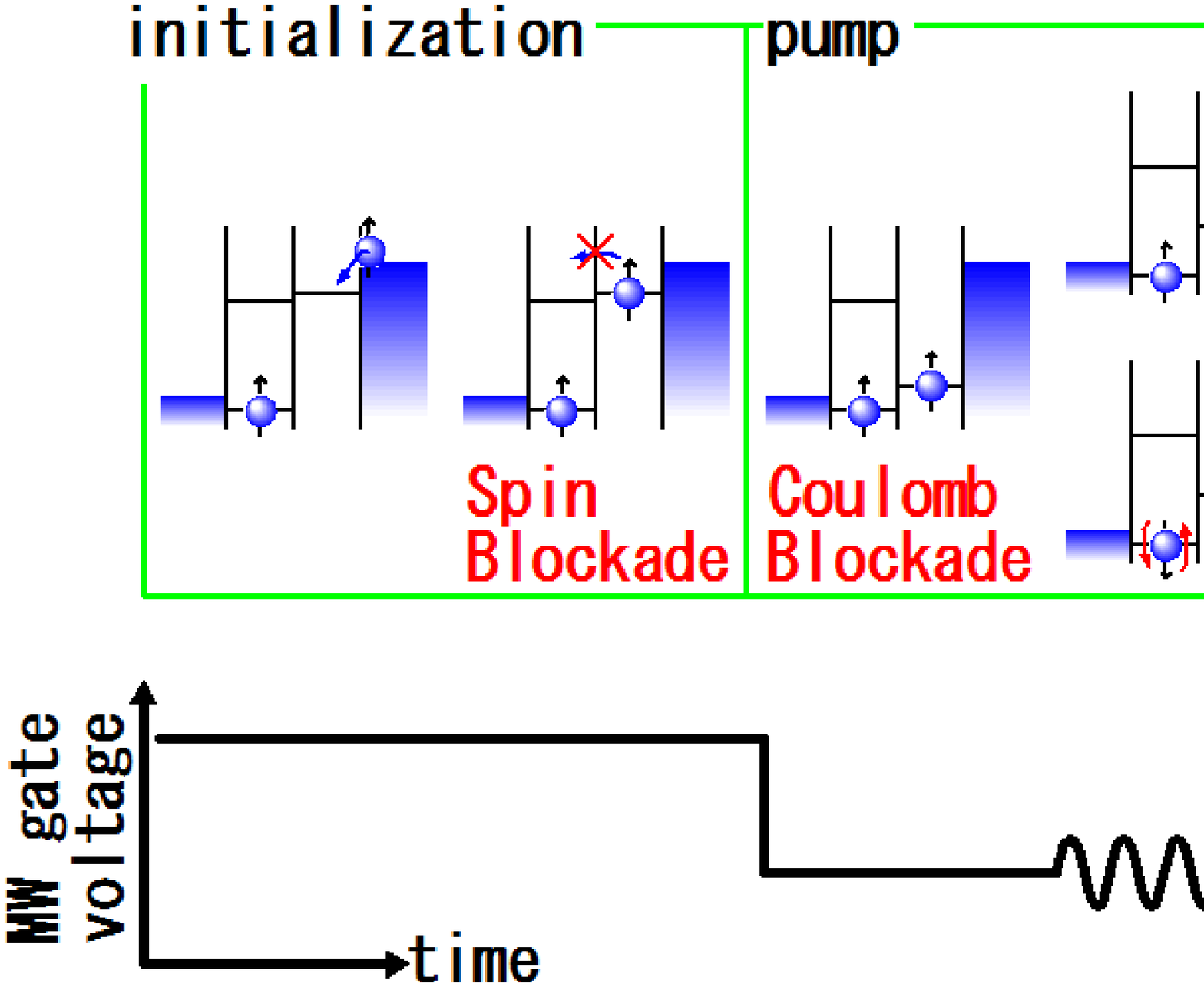} 
(a)\\
\end{flushleft}
\end{minipage}
\begin{minipage}{0.25\textwidth}
\begin{flushleft}
\includegraphics[width=\textwidth]{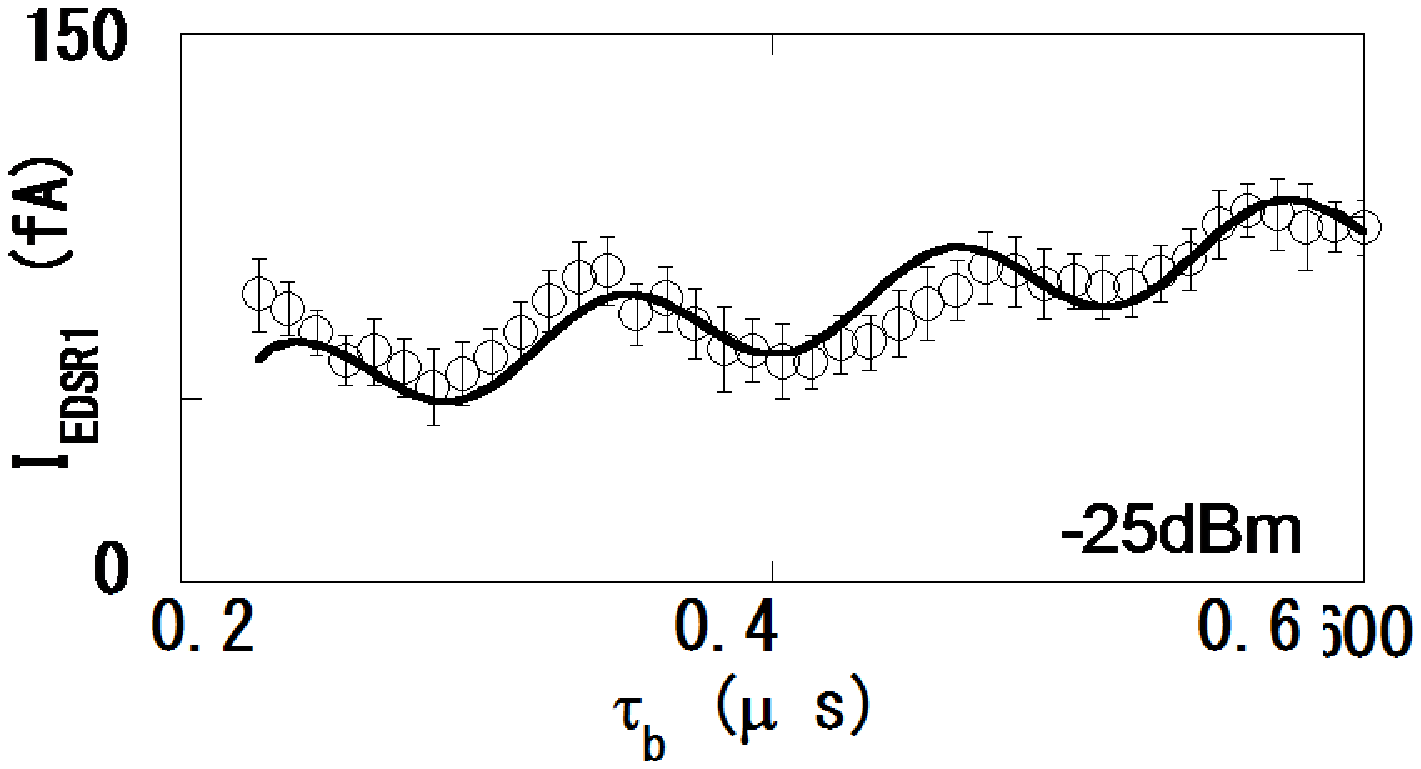}\\ 
(b)\\
\end{flushleft}
\end{minipage}
\\[5mm]
\begin{minipage}{0.225\textwidth}
\begin{flushleft}
\includegraphics[width=\textwidth]{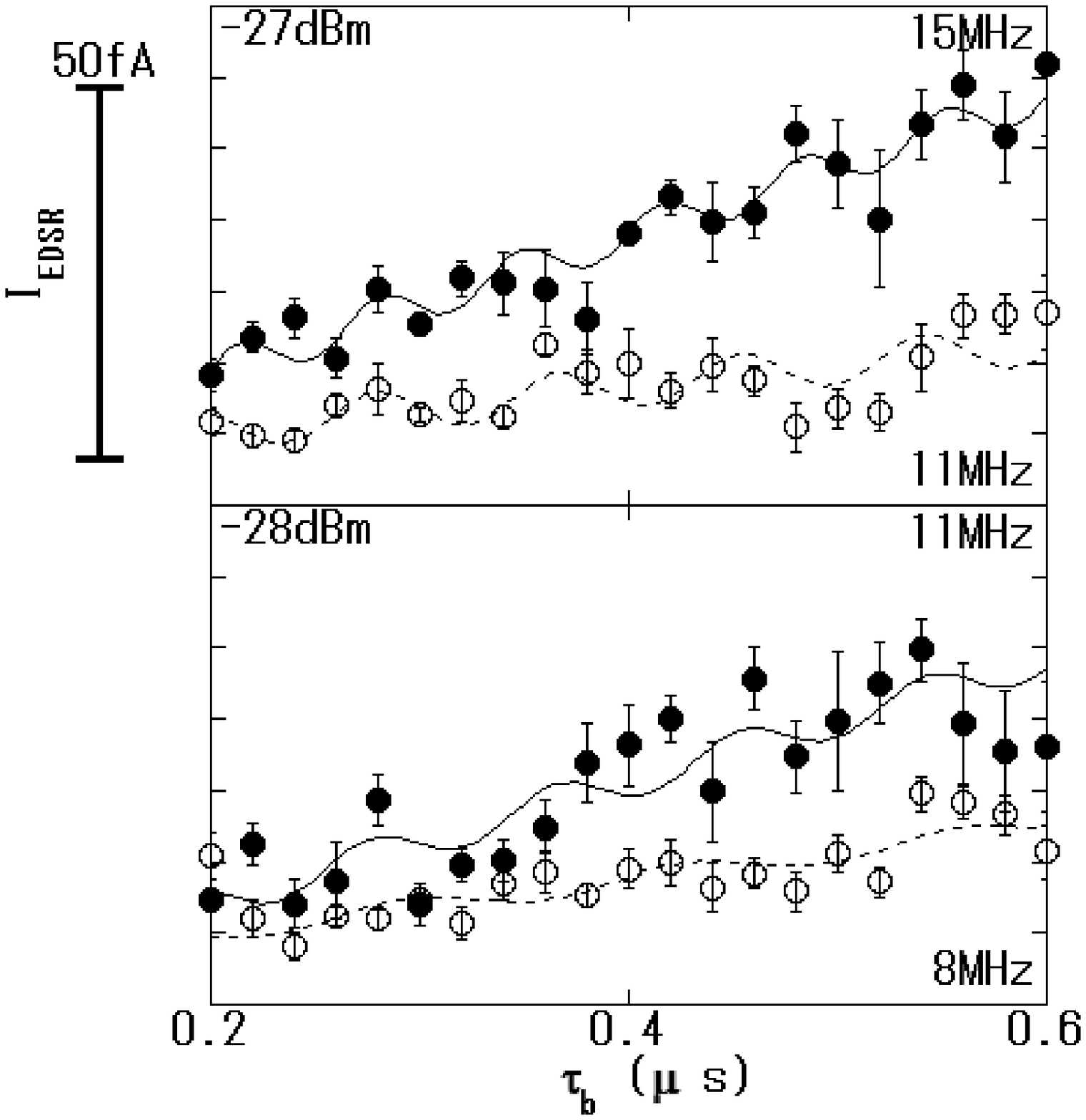} 
(c)\\
\end{flushleft}
\end{minipage}
\begin{minipage}{0.25\textwidth}
\begin{flushleft}
\includegraphics[width=\textwidth]{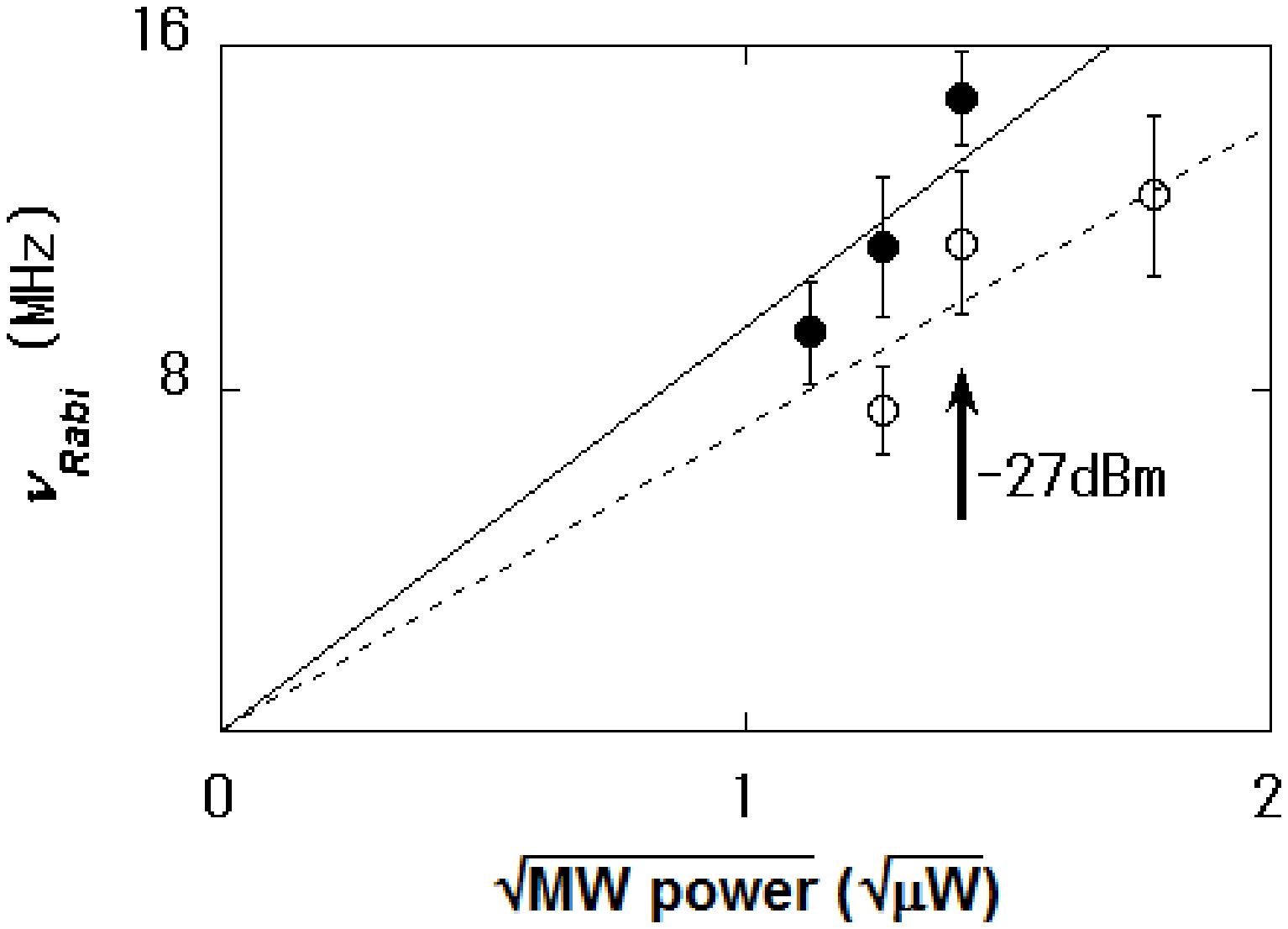} \\ 
(d)\\
\end{flushleft}
\end{minipage}
\\[5mm]
\begin{minipage}{0.30\textwidth}
\begin{flushleft}
\includegraphics[width=\textwidth]{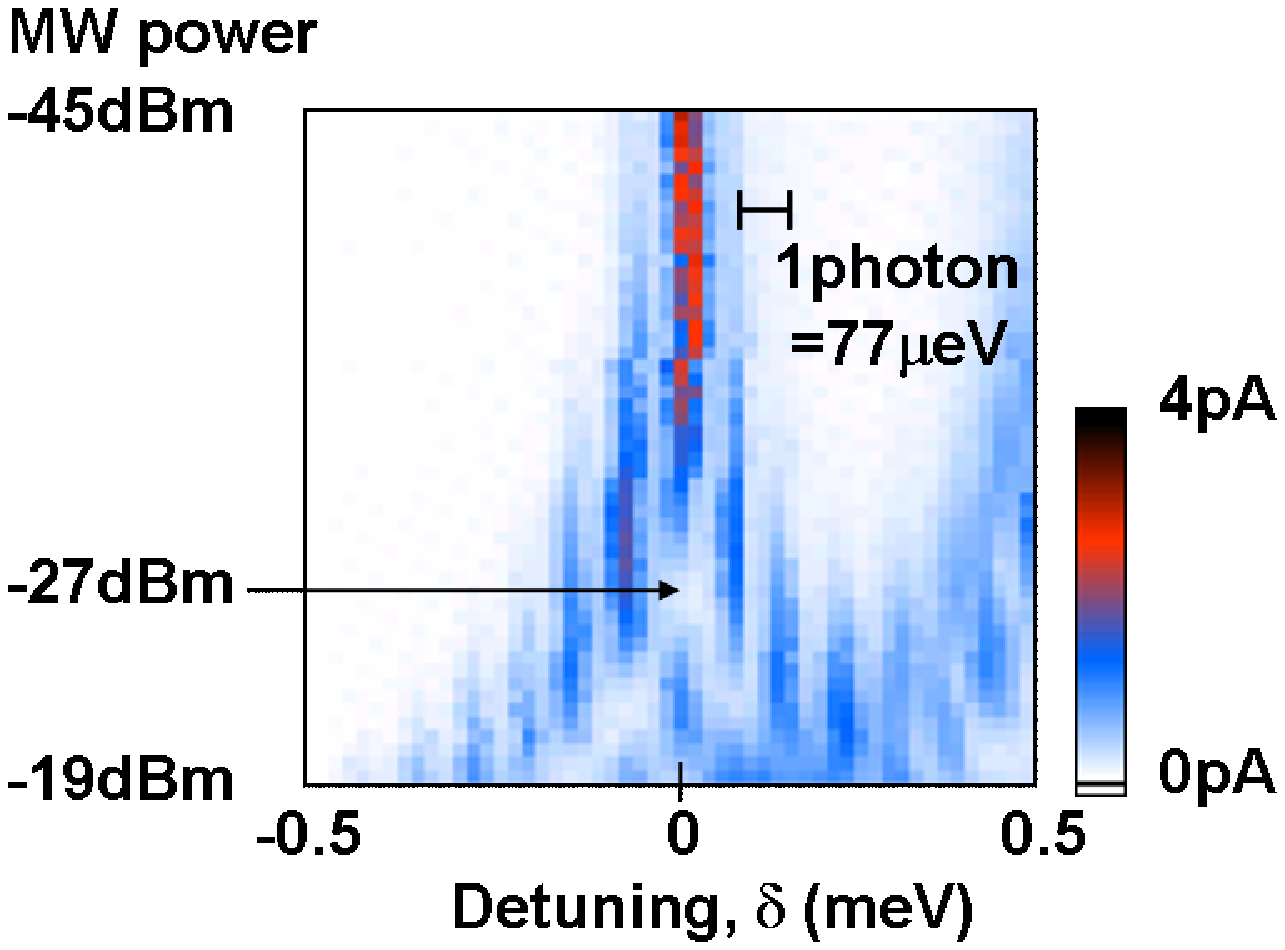} \\
(e)\\
\end{flushleft}
\end{minipage}
\caption{\label{fig:PP} 
(a) Schematic diagram of p-p measurement, including the MW gate voltage setting sequence. 
The pulse amplitude, repetition frequency, 
and MW frequency are 280 $\rm \mu$ eV, 500 kHz, and 18.5 GHz, respectively. 
(b) EDSR current vs MW burst time ($\tau_b$) measured for $I_{\rm EDSR1}$ (left dot). 
The solid line shows the fitting of a sinusoidal curve with a linear background (explained in the text) to Rabi oscillations of $I_{\rm EDSR1}$. (c) MW burst time dependence of the two EDSR peaks for two MW powers. 
The closed and open dots represent $I_{\rm EDSR1}$ and $I_{\rm EDSR2}$, respectively. 
The error bars represent the standard deviation of the current peak values of five measurements. 
(d) Rabi oscillation frequency, $\nu _{\rm Rabi}$, derived by fitting various MW $\nu_{\rm Rabi}$ values vs the square root of the MW power to the data in (c). The large error bars reflect the ambiguity in the sinusoidal fitting. 
(e) PAT data of $I_{\rm dot}$ vs inter-dot detuning $\delta$ measured for various powers. As the MW power is increased, more PAT peaks are observed. These PAT peaks are well reproduced by the square of Bessel functions with the inter-dot MW voltage drop as a fitting parameter.
}
\end{figure}

After the CW EDSR experiment, we readjusted the device conditions and performed a p-p experiment to observe Rabi oscillations. Here, we focus on two EDSR peaks observed at 3.30 and 3.32 T for an MW frequency of 18.5 GHz. 
The procedure of the p-p experiment is shown in Fig. \ref{fig:PP}(a) and consists of initialization, pump-and-probe stages. For initialization, the electron state is set in the P-SB condition where the configurations of ($\uparrow$,$\downarrow$), ($\downarrow$,$\uparrow$), and ($\frac{1}{\sqrt2}(\uparrow \downarrow- \downarrow \uparrow )$,0) are almost all degenerate as in Fig. \ref{fig:sampleSEM}(b). 
Hereafter, the arrows in the brackets show the electron states of each dot. $\uparrow$ and $\downarrow$ indicate the ground and excited spin states, respectively. The first arrow shows the electron state of the left dot and the second shows that of the right dot. $\frac{1}{\sqrt2}(\uparrow \downarrow- \downarrow \uparrow )$ is the spin singlet state in one dot and ``0'' indicates an empty dot. 
Under the P-SB condition, the two-electron state is initialized for one of the triplet states, for example, the ($\uparrow$,$\uparrow$) state. In the pump stage, the electron state is adiabatically moved to the region of the Coulomb blockade by tuning the MW gate voltage, and then an MW resonant with one of the electron spins is applied for a finite burst time, $\tau_b$. 
The electron spin flips coherently between the ground and excited states during the burst time.
This is Rabi oscillation. Finally, in the probe stage, we adiabatically restore the MW gate voltage to the initial P-SB condition after turning off the MW. 
If the spin is flipped in either dot, the hybridized singlet states of ($\uparrow$,$\downarrow$), ($\downarrow$,$\uparrow$), and ($\frac{1}{\sqrt2}(\uparrow \downarrow- \downarrow \uparrow )$,0) are readily formed to allow a finite leakage current.

We measured the EDSR current, $I_{\rm EDSR}$, for various $\tau_b$ values ranging from 600 to 200 ns in the p-p, as shown in Fig. \ref{fig:PP}(b). 
The p-p repetition rate was 500 kHz. Each data point is the average $I_{\rm EDSR}$ measured over 20 s. We had a technical problem with $\tau_b <$ 200 ns, and so we concentrated solely on $\tau_b >$ 200 ns. The oscillatory current with $\tau_b$ indicates Rabi oscillations with a frequency $\nu_{\rm Rabi}$ of 8.9 MHz in the sinusoidal curve fit including a linearly increasing background and a phase offset as $a + b \tau_b + c \sin ( 2 \pi \nu_{\rm Rabi} \tau_b + \phi_0)$. 
The current is proportional to the population of ($\frac{1}{\sqrt2}(\uparrow \downarrow- \downarrow \uparrow )$,0) just after switching to probe stage and we deduced that the population should be proportional to $\sin^2\left(2 \pi \nu _{\rm Rabi} \tau_b / 2 \right)=\frac{1}{2}\left(1-\sin\left(2 \pi \nu _{\rm Rabi} \tau_b \right)\right)$ by solving the master equation.
Here we neglect the damping effect. We explain the physical meanings of parameters $a$, $b$, $c$, and $\phi_0$ in the subsequent paragraphs. 
We also performed the same p-p measurement for the other EDSR peak at 3.32 T.
However, the p-p $I_{\rm EDSR}$ was too small to be resolved, probably because we had adjusted the device conditions to maximize the larger EDSR peak by sacrificing the smaller EDSR peak. 

We finally performed Rabi oscillation experiments for both spins to determine whether the Rabi 
frequencies are consistently characterized by the CW EDSR data with the MW power across each dot as a parameter. 
So, we readjusted several gate voltages to restore the $I_{\rm EDSR1}$ and $I_{\rm EDSR2}$ conditions of Fig. 2 to observe Rabi oscillations for both spins. 
We used the technique described in Ref. \onlinecite{Nowack} to average out the effects of nuclear spin polarization for the two peaks. For each fixed burst time, 
we recorded five current traces by externally sweeping the magnetic field $B_0$ five times. We swept the magnetic field from high to low to minimize the nuclear spin polarization effect.\cite{Jonathan,Vink} 
We observed large and small EDSR peaks, namely, $I_{\rm EDSR1}$ and $I_{\rm EDSR2}$, around 3.30 and 3.32 T, respectively, for each 20-h run of the $B_0$ sweep and then averaged $I_{\rm EDSR1}$ and $I_{\rm EDSR2}$. Figure \ref{fig:PP}(c) shows typical $I_{\rm EDSR}$ data vs $\tau_b$ measured at MW powers of -27 and -28 dBm in the upper and lower panels, respectively. We applied the same sinusoidal fitting as that in Fig. \ref{fig:PP}(b) to trace the oscillatory $I_{\rm EDSR}$ data. Although the data points are more scattered than those in Fig. \ref{fig:PP}(b)
, the $I_{\rm EDSR1}$ and $I_{\rm EDSR2}$ data sets 
are fairly well traced, using the common parameters $a=12.8 \pm 4.43$ fA and $c=2.8 \pm 0.66 {\rm fA}$ for both $I_{\rm EDSR1}$ and $I_{\rm EDSR2}$ and $b=8.36 \pm 0.83{\rm fA/\mu s \cdot mV}$ and $3.66 \pm 1.20 {\rm fA/\mu s \cdot mV}$ for $I_{\rm EDSR1}$ and $I_{\rm EDSR2}$, respectively. 
The phase offset $\phi_0$ is reported to be about $\pi/4$ reflecting the fluctuating nuclear field,\cite{Koppens,KoppensEcho} however, the value can be affected by the condition of the quantum dot.\cite{Rashba} 
Taking account of this point, we used $\phi_0$ as a fitting parameter but with a value between 0 and $\pi/4$ for each data set used in Fig. \ref{fig:PP}(b). 
Then, we managed to calculate $\nu_{\rm Rabi}$ values of 15 and 11 MHz for $I_{\rm EDSR1}$ and $I_{\rm EDSR2}$, respectively, as shown in the upper panel. We analyzed the data in the lower panel and calculated values of 11 and 8 MHz for $I_{\rm EDSR1}$ and $I_{\rm EDSR2}$, respectively.

Parameter $a$ is the current offset and is responsible for the positive probability of the probed singlet ($\frac{1}{\sqrt{2}}\left(\uparrow\downarrow-\downarrow\uparrow\right)$,0) state. This value is common to both oscillations as expected. Parameter $c$ is the Rabi oscillation amplitude and $b$ is the linear background with respect to $\tau_b$. 
We consider that EDSR is accompanied by a parallel leakage path of PAT between ($\uparrow$,$\downarrow$) or ($\downarrow$,$\uparrow$) and ($\frac{1}{\sqrt{2}}\left(\uparrow\downarrow-\downarrow\uparrow\right)$,0). 
In the pump stage, an electron spin turns downwards in a half period of Rabi oscillation. 
Before it turns upwards again, the electron can only move from the right dot to the left dot through PAT with multiple photons to compensate for the large voltage drop between the two dots. When PAT does not occur, the spin continues rotating in the right dot, but in the next Rabi cycle, the down spin has another chance of moving to the left dot by PAT. 
The probability of an electron moving is proportional to $\tau_b$, 
so the background leakage current increases linearly with $\tau_b$. It is also larger for larger EDSR peaks or faster Rabi oscillations, and we have just begun another experiment to enable us to understand the mechanism. 
The PAT process can cause extra decoherence; in the pump stage, the MW burst rotates one spin coherently, but the PAT can hybridize two spin singlet states before the MW burst finishes.

The correlation coefficients, $R^2$, evaluated for the sinusoidal fitting used here, are about 0.7 
while the value for linear fitting for the background is about 0.59. 
This is not very high and reflects the ambiguity caused by the scattering of the data points and the influence of a linearly increasing background. The data scattering is partly the result of the reduced averaging time for sweeping over the wider range of magnetic field, but it is mainly due to the fluctuating hyperfine field and spin resonance dragging effect.\cite{Koppens,Michel,Vink} These effects can be even more significant in the p-p measurement than in the CW ESR measurement 
because less averaging is employed.

The $\nu_{\rm Rabi}$ values obtained for different MW input powers are plotted in Fig. \ref{fig:PP}(d). 
The horizontal axis is the square root of the MW power, which is proportional to the driving voltage. 
All the data for each $I_{\rm EDSR}$ fall on a straight line crossing the origin. 
By comparing the slopes of the two straight lines, 
we find that $\nu_{\rm Rabi}$ for $I_{\rm EDSR1}$ for the left dot is 1.4 times larger than that for $I_{\rm EDSR2}$ for the right dot.

The Rabi frequency $\nu _{\rm Rabi}$ is proportional to the MW induced magnetic field, $B_{\rm MW}$, which is proportional to the MW induced electric field, $E_{\rm MW}$, across the double dot, i.e., $\nu _{\rm Rabi} = g \mu _e B_{\rm MW} / 2h = g \mu _e / 2h \times e E_{\rm MW} l_{\rm orb} ^2 / \Delta \cdot b_{\rm SL}$.\cite{Tokura,Michel} Here, the orbital spread, $l_{\rm orb}$, the QD confinement energy, $\Delta $, and the magnetic field gradient, $b_{\rm SL}$ are 48 nm, 0.5 meV, and 0.8 T/$\rm \mu$m, respectively.\cite{Michel} These are all fixed parameters in the present experiment. 
The only parameter that is varied is the MW electric field $E_{\rm MW}$ used to modulate the Rabi oscillation and therefore, the $\nu _{\rm Rabi}$ ratio between $I_{\rm EDSR1}$ and $I_{\rm EDSR2}$ is compared  directly with the $E_{\rm MW}$ ratio between the two dots.

For an MW power of -27 dBm, the mean value $\bar \nu _{\rm Rabi}$ between the two $\nu _{\rm Rabi}$ values is 13 MHz in Fig. \ref{fig:PP}(c). We use this frequency $\bar \nu _{\rm Rabi}$ to derive the mean values of $B_{\rm MW}$ and $E_{\rm MW}$ as $\bar B_{\rm MW}$ = 4.6 mT and $\bar E_{\rm MW}$ = 1.3 mV/$\rm \mu$m, respectively. 

\section{COMPREHENSIVE ANALYSIS OF CONTINUOUS WAVE AND PUMP-AND-PROBE EXPERIMENTAL RESULTS}

For a quantitative understanding of the $\nu_{\rm Rabi}$ ratio between the two dots, we estimated the $E_{\rm MW}$ distribution by measuring the inter-dot PAT. Under the P-SB condition in Fig. \ref{fig:sampleSEM}(b), there is actually another spin triplet state ($\uparrow \uparrow$,0), about 120 $\rm \mu$ eV above the ($\uparrow$,$\uparrow$) state. P-SB is not effective if the electrostatic potential of the left dot is reduced to achieve alignment between the two triplet states. 
We measured this triplet resonance current induced by PAT (Ref. \onlinecite{Nowack}) in Fig. \ref{fig:PP}(e). 
A fit to the theory provides a good estimation of the inter-dot voltage drop of 180 $\rm \mu$ V. 

The estimated distance from the MW gate edge to the center of the two QDs was 330 nm and the inter-dot distance was approximately 100 nm,\cite{Michel} with reference to the lithographic design in Fig. \ref{fig:sampleSEM}(a). The distances from the edge of the MW gate to the left and right dots were 280 and 380 nm, respectively. The MW gate width is much larger than the dot size, so we can assume that the MW electric field is uniform in the vertical direction in Fig. \ref{fig:sampleSEM}(a). We calculated the two-dimensional Coulomb potential using the above parameters to evaluate the spatial distribution of the MW electric field across the double dot. This calculation approach is valid because any surrounding metallic materials are thinner than the shielding length of a high frequency signal, which is called the skin depth. The electric fields evaluated at the left and right dots are 1.4 and 1.0 mV/$\rm \mu m$, respectively, for MW power of -27 dBm, and the average electric field matches the $\bar \nu _{\rm Rabi}$ value described above. The electric field at the left dot is 1.4 times larger than that at the right dot. This ratio is also comparable to the value of 1.58 calculated from the EDSR peak height in the CW measurement. 

\section{CONCLUSION}
We performed a continuous wave and pump-and-probe microwave experiment for a double quantum dot integrated with a micro-magnet for the selective observation of both the continuous wave electric dipole spin resonance current and the Rabi oscillation for a single electron in each dot. 
Both the ratios of the Rabi frequencies and the continuous wave electric dipole spin resonance currents between the two dots are consistently reproduced by the spatial distribution of the microwave electric field across the two dots. These results indicate that a double dot device with a micro-magnet will be useful for forming two or more spin qubits for quantum computing. 

\section*{ACKNOWLEDGMENTS}
We thank L. P. Kouwenhoven, D. Loss, L. M. K. Vandersypen, and R. Brunner for fruitful discussions. M.P.-L. acknowledges financial support from the Canadian Institute of Advanced Research. S.T. acknowledges financial support from the Grant-in-Aid for Scientific Research S (Grant No. 19104007) and B (Grant No. 18340081).
\newpage 

\end{document}